\documentstyle[12pt,a4,epsfig]{article} 

\title{ Higgs signals modified by singlet scalars } 
\author{ T.  Binoth, J.J. van der Bij }

\date{{\em Albert--Ludwigs--Universit\"at Freiburg, Fakult\"at f\"ur Physik \\
Hermann--Herder Str.3, D--79104  Freiburg}} 

\input epsf

\title{ Higgs signals modified by singlet scalars } 
\author{ T.  Binoth, J.J. van der Bij }

\date{{\em Albert--Ludwigs--Universit\"at Freiburg, Fakult\"at f\"ur Physik \\
Hermann--Herder Str.3, D--79104  Freiburg}} 

\begin{document}

\maketitle

\vspace{1.5cm}

\hspace{9cm}  Freiburg--THEP--94/26. 

\hspace{9cm}  hep--ph/9409332. 

\vspace{1.5cm}

\begin{center} Presented by T. Binoth at the international Seminar ''Quarks 94''  \\
                     in Vladimir CIS  from May 11 to May 18. 
\end{center}

\vspace{1.5cm}

\abstract{\em The influence of massless scalar singlets on Higgs signals is discussed.
It is shown that in the case of strong interactions between the Higgs boson and the singlet 
fields, signals could be suppressed in such a way that detection of the Higgs boson becomes 
impossible.} 

\newpage

\section{Introduction}
After the recent Fermilab results the particle content of the Standard Model (SM) is 
confirmed with the exception of the Higgs boson. 
 Because radiative corrections due to the Higgs sector
 only give rise to  terms logarithmic in the Higgs mass, there are no reasonable experimental 
 upper bounds for its mass. For the direct search at hadron colliders 
 different Higgs signals are 
 discussed in the literature \cite{Bijetal}. As long as $M_{Higgs} < 2\, M_Z$ these signals are 
 difficult to detect because of the large backgrounds. 
 Beyond the vector boson threshold one expects rather clean signals of double leptonic decays 
 of Z--boson pairs $H\rightarrow ZZ \rightarrow l^+l^-l^+l^-$. 
 Thus, at the LHC there is good hope
 to find the Higgs boson of the SM, as long as $2m_Z<m_H < 700 GeV$. The most effective production
 mechanism in this energy range is the gluon fusion process:
\begin{figure}[h]
\begin{center} 
Signal: $gg\rightarrow \sigma \rightarrow ZZ \rightarrow l\bar l l \bar l$\\
Background: $q\bar q \rightarrow ZZ$\\
Background (box): $gg \rightarrow ZZ$ \\
\parbox{5in}{\caption{\em Higgs boson production by the gluon fusion process with subsequent
decay into four leptons. The background reactions are shown, too.}}
\end{center}
\end{figure}
The calculation of this process predicts a reasonable 
 number of such events. 
 (For $m_H=200\,GeV,m_t=150\,GeV$ and a luminosity at LHC of $\sim 170/fb/year$ 
 one can estimate a few hundred events per year).

The question arises, how such signals are changed by nonstandard particles.
Every additional non standard decay channel of the Higgs boson leads to a reduction of the signal.
An example of non standard particles are the neutralinos of the supersymmetric version of the SM.
In this talk, I will consider the simplest way to extend the SM by unobservable particles.
I just add scalar singlet fields -- which I call  phions during this talk -- to the SM. Because they are 
assumed to be singlets under the SM group $SU(3)_C\times SU(2)_L\times U(1)_Y$, 
there are no couplings to any known fermion. The phion can only interact with the 
Higgs boson. This kind of scalars occur for example 
in non minimal extensions of the supersymmetric SM. If now $2m_{\varphi}<m_H$, Higgs decay into phions is 
not only possible, but in the case of strong coupling it is the most important decay channel, leading
to significant experimental consequences. 

After this introduction, I want to present a model of scalar singlet fields coupled, not necessarily
weakly, to the SM Higgs boson. Afterwards the influence of the phions on Higgs signals like $gg\rightarrow H
\rightarrow ZZ \rightarrow 4l $ is discussed in detail.  
   
\section{The extended scalar sector}

{\it 2.1 The model}

\smallskip

To the usual SM Higgs sector we just add a singlet scalar sector, given by a $\varphi^4$ theory. 
Renormalizability allows for one nontrivial interaction term. 
To deal with the strong
interactions we introduce an $N$--plet of 
 phions, then we are able to use the nonperturbative $1/N$--expansion. \\
The model neglects gauge interactions and Yukawa couplings of the SM. It is:
\begin{eqnarray} \label{1model}
L_{Scalar} &=& L_{Higgs} + L_{Phion} + L_{Interaction}   \qquad
 \mbox{where}  \nonumber \\
 L_{Higgs}  &=& 
 - \partial_{\mu}\phi^+ \partial^{\mu}\phi -\lambda_0 (\phi^+\phi - v^2/2)^2 \nonumber \\
  L_{Phion}  &=& - 1/2\,\partial_{\mu} \vec\varphi^T \partial^{\mu}\vec\varphi
     -1/2 \, m_{\varphi\,0}^2 \vec\varphi^2 - \kappa_0/(8N) (\vec\varphi^2 )^2 \nonumber \\
  L_{Inter.}  &=& -\rho_0/(2\sqrt{N})\, \phi^+\phi\, \vec\varphi^2 
\end{eqnarray}  
Here we use a metric with signature $(-+++)$. 
$\phi=1/{\sqrt{2}}(\sigma+v+ig_1,g_2+ig_3)$ is the SM complex Higgs doublet 
with 
$<0|\phi|0> = (v/\sqrt{2},0)$, where $\sigma$ is the physical  Higgs boson and $g_{i=1,2,3}$ 
are the three Goldstone bosons. $\vec\varphi = (\varphi_1,\dots,\varphi_N)$ is a real vector with 
$<0|\vec\varphi|0>= \vec 0$. $\lambda_0,\rho_0,\kappa_0,m_{\varphi\,0}$ are bare parameters.
In the unitary gauge operators of the form
\begin{center}
{\em Picture 1}
\end{center}
are introduced by loop corrections. Assuming strong interactions in the extended scalar sector
($\rho$, $\kappa$ nonperturbative) such loop effects are not suppressed. To deal with 
such nonperturbative effects we use the $1/N$--expansion \cite{Ein}. 
\newpage

\noindent
{\it 2.2 $\quad 1/N$--expansion}

\smallskip

We are only interested in operators with external Higgs legs. These we can classify in types,
 (a) with and (b) without internal Higgs lines. Diagrammatically:
\begin{center}
{\em Picture 2}
\end{center}
The former with $k$ legs are got just by closing two lines of a $k+2$ operator, so its enough 
to focus on the later, for reasons which become clear below. These we can sort with 
respect to the Higgs legs ($k$) and powers ($n$)
of $N$. Every $\rho$--vertex counts $n=-1/2$, every $\kappa$--vertex counts $n=-1$ and every
 phion loop counts $n=1$. The highest $n$ operators have $n=1/2$. They are 
\begin{center}
{\em Picture 3}
\end{center}
Both contributions to the Higgs propagator are constant terms for any fixed value of $N$. 
The tadpole contribution is taken
 into account by using the experimental value of the vacuum expectation value (vev) 
 $v = 246\,GeV$, the
 constant selfenergy term can 
 be absorbed by a bare mass term for the complex Higgs doublet. The $n=0$ operators are 
 an infinite sum of loop graphs, a so called bubble sum. They are \\
\begin{center}
{\em Picture 4}
\end{center} 
No other structures with $n=0$ are possible. Operators with higher $n$ will be suppressed 
by factors of $(1/N)^n$, which can be made arbitrarily small by taking $N$ sufficiently large. 
We only want to discuss this large--$N$ case, the formal limit $N\rightarrow \infty$. 
The upper $n=0,\, k=3,4$ bubble sums are leading to a renormalization of
the Higgs coupling $\lambda$, which we want to keep perturbatively. (Strictly speaking one has to
check that the quartic Higgs coupling stays in the perturbative region for a given energy range 
even if the other couplings ($\rho,\kappa$) are nonperturbative. This can be shown by 
renormalization group analysis of the model (Eq.\ref{1model}).) Operators which are 
built out of these 4 Higgs vertices by closing two lines
are then suppressed by powers of $\lambda$. 
This is the reason for neglecting them from the beginning. 
 
The bubble sum in the 
Higgs propagator will modify the position of the pole and so the mass and the width of the 
Higgs boson. The $m$--th term of the sum is
\begin{center} 
{\em Picture 5} $ = v^2 \rho_0^2 \, I \,(-\kappa_0\,I)^{m-1}$ 
\end{center}
which leads to a geometric series
\begin{eqnarray} \label{2self}
\Sigma(p^2,m_{\varphi}^2) = -v^2\rho_0^2I \, \sum \limits_{m=0}^{\infty}(-\kappa_0 I)^m
                      = -\frac{v^2\rho_0^2I}{1+ \kappa_0I}\quad .
                      \end{eqnarray}      
In the upper expressions $I=I(p^2,m_{\varphi}^2)$ is the euclidean integral
\begin{displaymath}
I(p^2,m_{\varphi}^2) := \frac{1}{2} \int \frac{d^4k}{(2\pi)^4}\,
\frac{1}{k^2+m_{\varphi}^2}\,\frac{1}{(k+p)^2+m_{\varphi}^2}\quad .
\end{displaymath}     
The infinities of the theory can be absorbed in the bare quantities \cite{CJP}. After the 
renormalization 
procedure one is left with renormalized couplings ($\rho$, $\kappa$) depending on a renormalization
scale, say $\mu$. The infinity of the integral is then canceled and we arrive at the regular
expression 
\begin{eqnarray} \label{3self}
\Sigma^{reg}(p^2,m_{\varphi}^2,\mu^2) 
= -\frac{v^2\rho^2(\mu^2) I^{reg}(p^2,m_{\varphi}^2,\mu^2)}{1+ \kappa(\mu^2) 
I^{reg}(p^2,m_{\varphi}^2,\mu^2)}  \qquad ,
\end{eqnarray}
where
\begin{eqnarray} 
I^{reg}(p^2,m_{\varphi}^2,\mu^2) &=& I(p^2,m_{\varphi}^2) - I(0,\mu^2) \nonumber \\
 &=& \frac{1}{32\pi^2} (  \log(\frac{\mu^2}{m^2_{\varphi}}) +2 -
  \sqrt{1+\frac{4m_{\varphi}^2}{p^2}} 
  \log \frac{\sqrt{1+\frac{4m_{\varphi}^2}{p^2}} +1 }{ \sqrt{1+\frac{4m_{\varphi}^2}{p^2}}-1} )
  \nonumber                        
\end{eqnarray}         
We want to discuss the case where the center of mass energy flowing through the graph 
 is far beyond the production threshold of the phions, $s=-p^2>>4m^2_{\varphi}$. There, 
 the phions will considerably influence the Higgs signals because the decay of the Higgs boson  
 into phions is possible. In that  region the integral is independent of the phion mass or in other 
 words we calculate the limit of massless phions. To avoid additive constants  we rescale the 
 renormalization scale $\mu \rightarrow \mu/e$. Then
\begin{equation} \label{4limint}
I^{reg}(-s,m_{\varphi}^2,\mu^2) = -\frac{1}{32\pi^2} \log(\frac{-s}{\mu^2})
\, , \quad s >> 4m^2_{\varphi} \quad .
\end{equation}      
We are now ready  to write down the inverse Higgs propagator. Putting $v=1$, means all energies and 
masses are given in units of $v=246\,GeV$, it is
\begin{eqnarray} \label{5prop}
D_{H}^{-1}(s,\mu^2) &=& -s + 2\lambda + \Sigma^{reg}(s,\mu^2) \nonumber \\
\Sigma^{reg}(s,\mu^2) &=&
 \frac{\rho^2/(32\pi^2)\log(-s/\mu^2)}{1-\kappa/(32\pi^2)\log(-s/\mu^2)} 
\end{eqnarray}
The couplings $\lambda,\kappa,\rho$ are renormalization scale dependent quantities. 
Because they are not fixed at any scale we can treat them as free parameters in the following.
Note that because of the negative argument of the logarithm the selfenergy $\Sigma$ 
is a complex number. 

For the denominator of the selfenergy a comment is in order. 
Beyond the Landau pole of the $\kappa/N\varphi^4$ theory ($s\sim \mu^2\exp(32\pi^2/\kappa)$) 
the real part of the Higgs selfenergy behaves like $ \sim -\rho^2/\kappa $. 
This could lead to a negative mass squared for the Higgs propagator. On the other hand the 
Landau pole would lead to the breakdown of the $1/N$--expansion even for the pure 
$\kappa/N\varphi^4$ theory \cite{Root}. This would be no problem as long as 
$\kappa/32\pi^2$ is a small number. Then the pole would occur beyond observable energies. 
As was shown in
the literature \cite{KoKu,AKS}, the problem is cured, if one allows for negative values of 
$\kappa$. In the limit $N\rightarrow \infty$ the classical theory makes no difference between 
the sign of $\kappa$. One  always arrives at a non interacting theory. But at the quantum 
level one finds a stable theory only for 
the negative sign. Then no Landau pole occurs. 
 A spontaneously broken $O(N)-\kappa\varphi^4$ theory leads always to Landau poles which causes 
 ill defined propagators in the case of strong
  interaction. For this reason, the $1/N$--expansion used for a strongly interacting 
   Higgs sector hardly produces meaningful results. We  avoided too  big a 
   Higgs coupling $\lambda$ too, because we do not know how to treat it correctly.
  Using a negative quartic phion coupling we find an expression 
  for the Higgs propagator, which is the starting point for the following analysis. 
\begin{eqnarray} \label{6higgsprop}
 D_{H}^{-1}(s,\mu^2) &=& -s + 2\lambda + 
       \frac{\alpha\log(-s/\mu^2)}{1+\beta\log(-s/\mu^2)} \quad ,
\end{eqnarray}
with the abbreviations 
  $\alpha=\rho^2/(32\pi^2)$ and $\beta = |\kappa|/(32\pi^2)$.
Later we will see, that the model is interesting in the region, where $\beta$ is small and no 
problems with the Landau pole occur.  

\medskip

\noindent
{\it 2.3 Mass and width effects due to the hidden phions} 

\smallskip

We are now able to fix the physical observables of the Higgs boson, its mass $M$ and width 
$\Gamma$ in terms of $\lambda,\alpha$ and $\beta$. Note that we still neglect other SM particles.
To this end, we have to find the pole of the Higgs propagator (Eq.\ref{6higgsprop}). 
This is done by putting \begin{displaymath}
s= -p^2=(M-i\Gamma/2)^2=s_0\exp(-i2\theta), \quad s_0:=M^2+\Gamma^2/4 \end{displaymath}
into the inverse propagator  and searching for the zeros of the real and imaginary part. 
We use the convenient renormalization scale $\mu^2=s_0$ which provides us with the simple 
substitution $\log (-s/\mu^2) \rightarrow -i(2\theta +\pi)$. It follows
\begin{eqnarray} \label{pole}
s_0 \cos 2\theta &=& 2\lambda + \frac{\alpha\beta(2\theta+\pi)^2}{1+\beta^2(2\theta+\pi)^2} 
\nonumber\\
s_0 \sin 2\theta &=& \frac{\alpha(2\theta+\pi)}{1+\beta^2(2\theta+\pi)^2}\nonumber \\
M = \sqrt{s_0} \cos \theta & , & \Gamma =2 \sqrt{s_0} \sin \theta =2M\tan\theta\quad .
\end{eqnarray}
After eliminating $s_0$ one finds numerically the angle $\theta(\lambda,\alpha,\beta)$. 
In the limits ($\alpha\rightarrow 0,\beta$ fix) and ($\beta\rightarrow \infty,\alpha$ fix) the 
corrections to the Higgs mass and width are of order $\alpha,1/\beta$ respectively. The model gets 
interesting 
if $\alpha,\beta \sim 1$. This is shown in figure (2a,b). 
\begin{figure}
\begin{center}
\setlength{\unitlength}{1.0cm}
\begin{picture}(14.0,19.0)
\put(0.5,18){{\Large Figure 2a)}}
\put(0.5,8){{\Large Figure 2b)}}
\put(4,17){$\alpha$}
\put(4,7){$\alpha$}
\put(7,10.5){$\beta$}
\put(7,0.5){$\beta$} 
\put(0.5,14){$\frac{m_H}{GeV}$} 
\put(0.5,4){$\frac{\Gamma_H}{GeV}$}

\put(-2.0,-3.0){
                \epsfxsize14.0cm
                \epsfysize14.0cm
                \epsffile{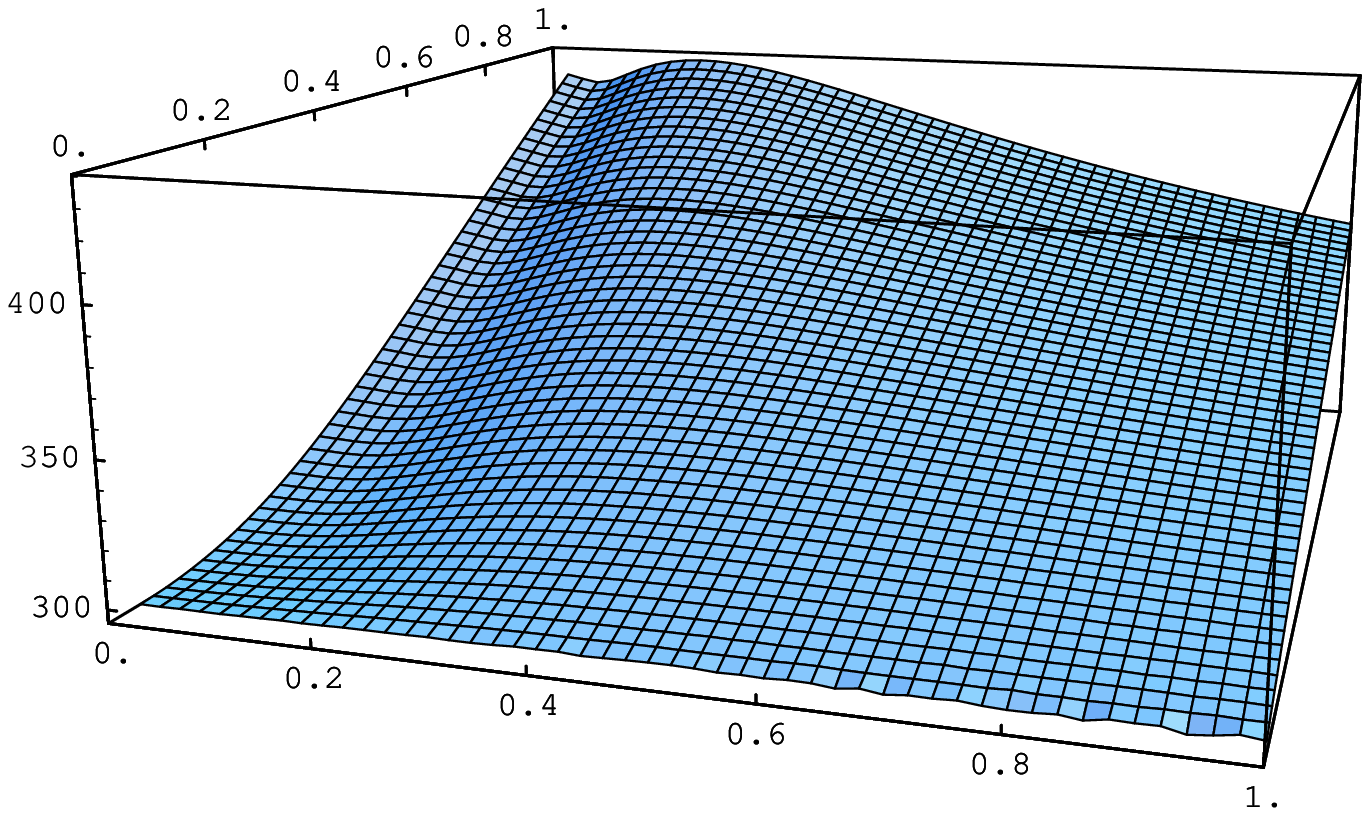} }
\put(-1.5,-13.0){
                \epsfxsize14.0cm
                \epsfysize14.0cm
                \epsffile{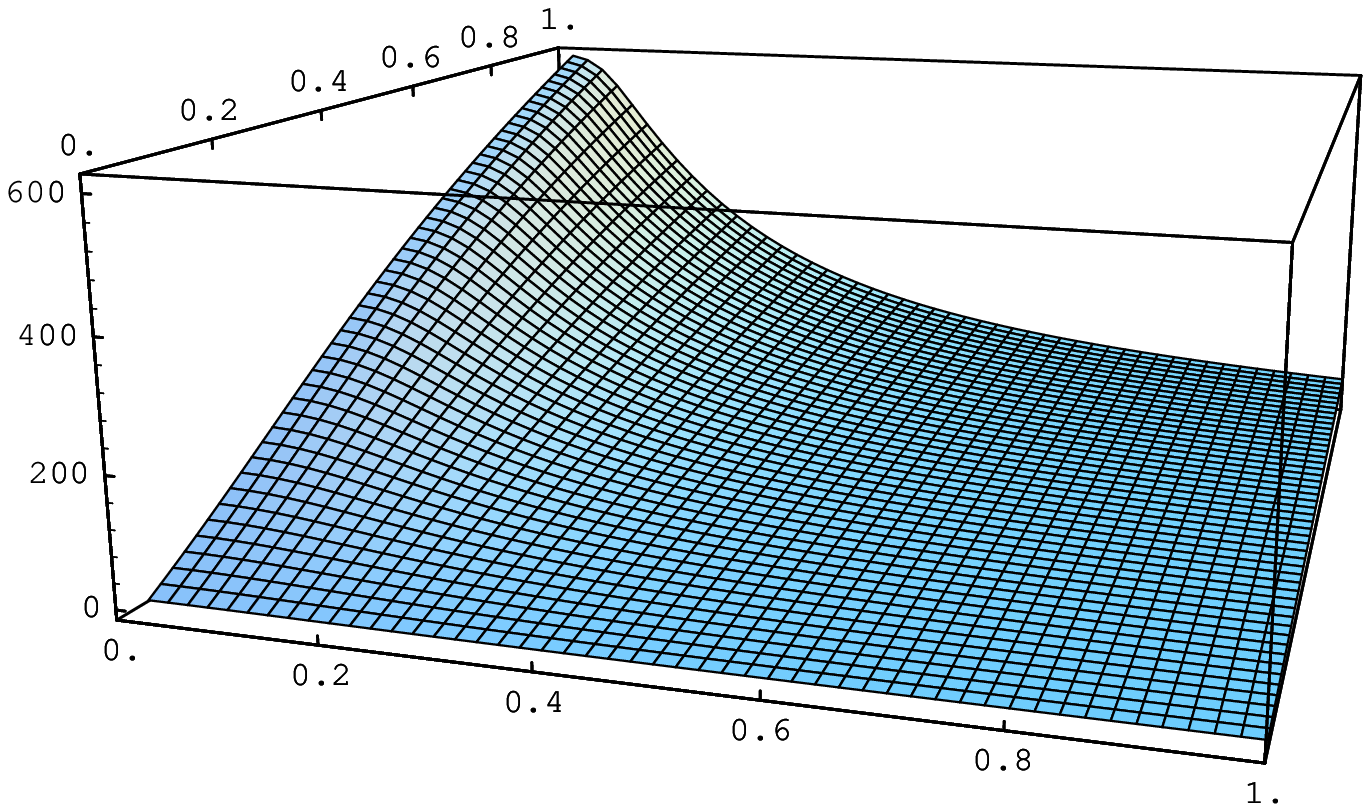} }
\end{picture}
\parbox{5in}{\caption{\em (a) Behaviour of the Higgs mass in 
 the $\alpha$, $\beta$ plane for  
 $\lambda=0.744$.  (b) The same as in (a) for the decay width.}}
\end{center}
\end{figure}
One finds an increase of the Higgs width with growing $\alpha$ and a decrease in the $\beta$ 
direction. 
The $\alpha$ dependence is due to the enhanced decay probability, the $\beta$ dependence 
indicates that 
strong interactions in the phion sector suppress the decay into phions somewhat, in the case of positive
$\beta$ (negative phion self coupling).
 Figures (2a,b) indicate that the most drastic consequences for  Higgs resonance signals 
 will occur in the 
 region where $\beta<<1$ and $\alpha>>\beta$. There, the ratio $\Gamma/M$ gets bigger than one 
 leading to unobservable broad resonances. 

In the following we want to give the results for the case $\beta=0$. Then Eq.(\ref{pole}) 
simplifies and the functions $M,\Gamma$ over $\alpha$ are easily found. The limiting cases 
are given in the  following table.
\begin{table}[h]
\begin{center}
\begin{tabular}{|c|c|c|c|c|c|}
\hline 

$\alpha$ & $\theta$ & $s_0$ & $M$ & $\Gamma$ & $\Gamma/M$ \\
\hline
$\rightarrow 0$ & $\sim\frac{\alpha\pi}{4\lambda}$ & $\rightarrow 2\lambda$ & 
$ \rightarrow\sqrt{2\lambda}$
& $\sim\frac{\alpha\pi}{\sqrt{2\lambda}}$ & $\frac{\alpha\pi}{2\lambda}$ \\
$\rightarrow\infty$ & $\rightarrow \frac{\pi}{4}$ & $\sim\frac{3\pi}{2}\alpha$ & 
$\sim\frac{1}{2}\sqrt{3\pi\alpha} $ & $\sim \sqrt{3\pi\alpha}$ & $\rightarrow 2$ \\ 
\hline  
\end{tabular}
\parbox{5in}{\caption{\em Higgs mass, width in the case $\beta=0,\alpha\rightarrow 0,\infty$.}}
\end{center}
\end{table} 
In figure (3) the mass and 
the width is plotted for four different values of $\lambda$ which belong to the treelevel 
Higgs masses of 100, 200, 300 and 400 GeV. These plots indicate an unbounded growth of 
the Higgs mass and width 
as  $\alpha$ increases. 
\begin{figure}
\begin{center}
\setlength{\unitlength}{1.0cm}
  \begin{picture}(14.0,15.0)
  \put(1.5,-1.0){
    \epsfxsize12.0cm
    \epsfysize18.0cm
    \epsffile{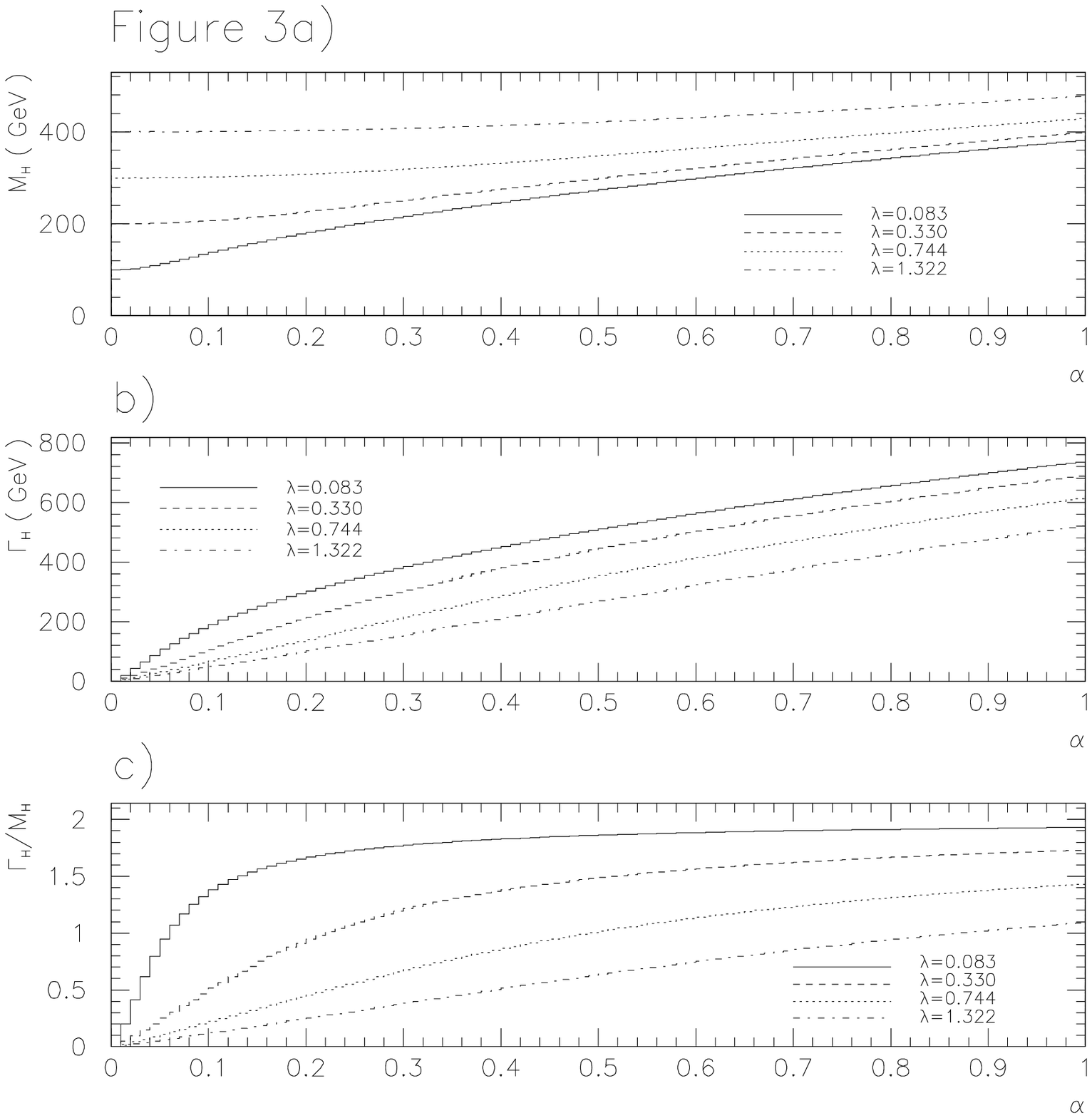} }
  \end{picture}

\parbox{5in}{\caption{\em
(a): Higgs mass vs. $\alpha$ for several values of $\lambda$ with $\beta=0$.
(b): Higgs width vs. $\alpha$.
(c): Width to mass ratio vs. $\alpha$. }} 
\end{center}
\end{figure}
   
For Higgs signals the ratio $\Gamma/M$ is most important. We plot this ratio as a function of $\alpha$
in figure (2c). The ratio  asymptotically approximates the value two.
 The bigger the value of $\lambda$, or the SM Higgs mass, the 
slower the ratio reaches its asymptotic value. The denominator is simply bigger in this case. 
The physical consequence of all that is that every Higgs resonance will be washed out. No sharp 
signal would occur. 
The strong, invisible decay into phions will suppress all signals which depend on the 
Higgs decay products.   
Note that we have calculated everything in the limit of massless phions (Eq. \ref{4limint}). 
A more detailed analysis of mass effects is deferred to  a future paper.  

To be more explicit the influence of the hidden sector to the hopeful leptonic $H\rightarrow ZZ$ 
decay is discussed in the next section.  

\section{The signal $H\rightarrow ZZ \rightarrow 4\mu$ }

At hadron colliders with beam energies in the $TeV$ range like the LHC, Higgs bosons 
can be produced by the gluon fusion process. As long as $m_H>2m_Z$ the subsequent decay
of the $Z$--pair into two lepton pairs provides us with a very clear signal in the case of a SM 
 Higgs boson \cite{Bijetal}. To see how our model (Eq. \ref{1model}) modifies the SM results one has 
to calculate the $gg\rightarrow ZZ\rightarrow 4l$ process with the Higgs propagator modified as  in 
Eq.(\ref{6higgsprop}). In addition one has to include the SM Higgs width due to possible  
decays into SM particles 
like the massive vector bosons and the Top quark. 
Again we have put $\beta=0$ because we are interested in a worst case scenario. 
The leptons are muons in our calculation.
First we present a table with the  cross sections  for different values of $\alpha$ and 
$\lambda$ computed for the LHC energy of $\sqrt{s}=16$ TeV.
We have subtracted the background due to the $q\bar q\rightarrow ZZ$ and the $gg\rightarrow ZZ$ 
box graph. 
\begin{table}[h] 
\begin{center}
\begin{tabular}{|c|cccccc|}
\hline
$\sigma$-background (fb), $\alpha=$ & 
          $0.0$ & $0.2$ & $0.4$ & $0.6$ & $0.8$ & $1.0$ \\
\hline 
$\lambda=0.33$ & 2.98 & 0.08 & 0.04 & 0.02 & 0.01 & $<$0.01 \\
$\lambda=0.52$ & 2.80 & 0.13 & 0.06 & 0.03 & 0.02 & 0.01 \\
$\lambda=0.74$ & 3.10 & 0.43 & 0.11 & 0.06 & 0.04 & 0.02\\
$\lambda=1.32$ & 2.08 & 0.74 & 0.22 & 0.14 & 0.09 & 0.06\\
\hline
\end{tabular}
\parbox{5in}{\caption{\em Cross section minus background for $pp \rightarrow 4\mu+X$ 
at LHC for different values of $\alpha$ and  $\lambda$. }}
\end{center}
\end{table}

The $\lambda$ values in table (2) correspond
to SM Higgs boson masses of $200$, $250$, $300$ and $400$ GeV respectivly. The Top mass is assumed to be $150\,GeV$ which is the reason for 
the enhancement of the cross section for a Higgs boson of mass $300\,GeV$ ($\lambda=0.744$).
To take into 
account the acceptance of a detector 
we placed a rapidity cut of $|y_l|<3$ and a transverse momentum cut of $p_{Tl}>20\,GeV$ on the 
muons.   
As an example we have plot  the 
invariant mass spectrum of the outgoing $Z$--pair in 
figure (4) for $\lambda=0.744,\,(m_{Higgs}=300\,GeV)$.
\begin{figure}
\begin{center}

\setlength{\unitlength}{1.0cm}
  \begin{picture}(14.0,17.0)
  \put(1.5,-1.0){
    \epsfxsize12.0cm
    \epsfysize18.0cm
    \epsffile{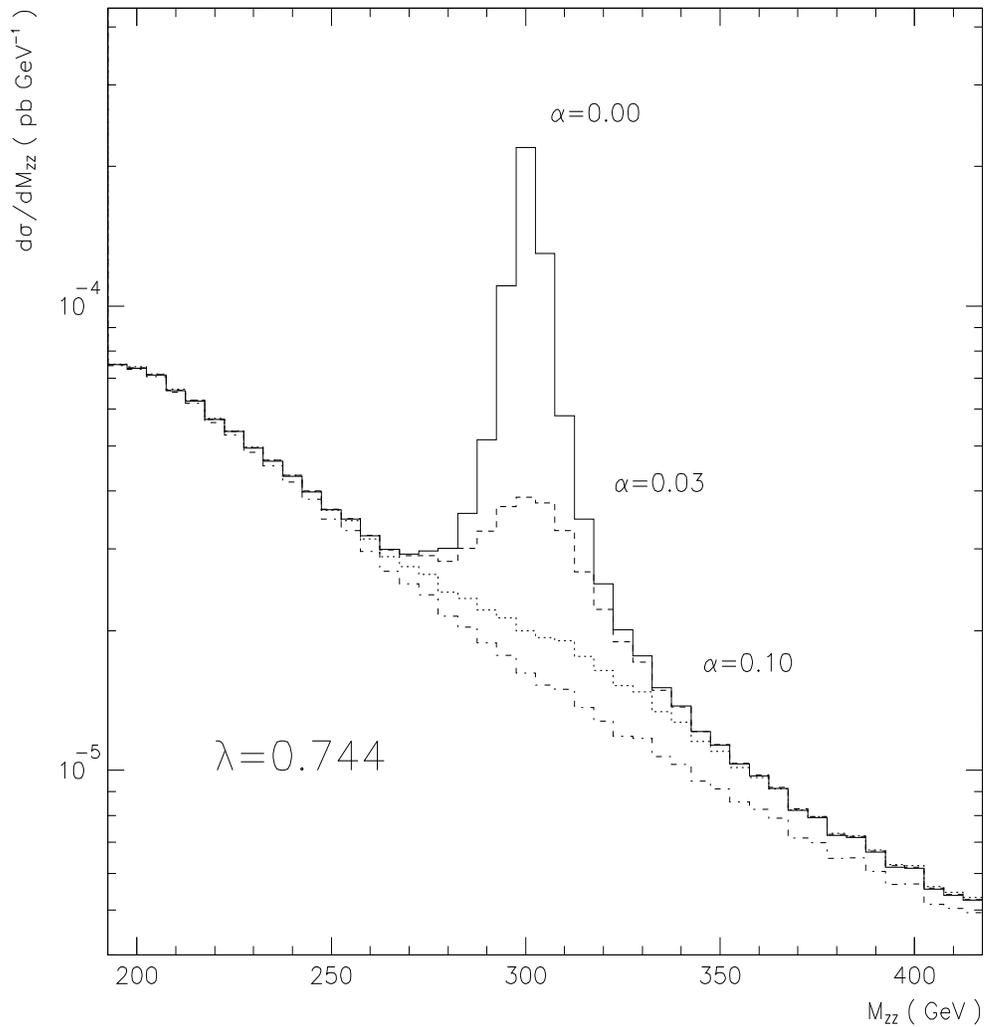} }
  \end{picture}

\parbox{5in}{\caption{\em Invariant mass distribution for  $Z$--pairs at the LHC. 
We assumed a rapidity 
cut of $|y_l|<3$  and a transverse momentum cut of $p_{Tl}>20\,GeV$ for the outgoing muons. }}
\end{center}
\end{figure}
The slight 
shift of the pole is the $\alpha$ dependent mass renormalization. For small $\alpha$ a weak signal
is  still  visible. As $\alpha$ gets larger no signal survives. This is due to the  increase
of the Higgs width and, at the same time, the supression of the branching ratio 
$R=\sigma(H\rightarrow ZZ)/\sigma(H\rightarrow ZZ,WW,t\bar t,\varphi\varphi )$, as $\alpha$ gets 
a strong coupling. 

\smallskip
  
For different values of $\lambda$  table (2) and figure (4) clearly reflect the arguing of 
the last section (2). Due to the preferred strong decay into phions the $ZZ$ signal is considerably 
suppressed. The width of the peak increases so drastic that the signal ceases to dominate 
over the background. The decrease of the signal with growing $\alpha$ is faster 
for smaller $\lambda$ (lighter Higgs) which is a consequence of the smaller width to mass ratio.  

For Higgs signals with a Higgs mass below the $Z$--pair threshold the situation is similar. Every 
signal which depends on the decay products of the Higgs boson are significantly suppressed by
strong decay into invisible particles. 

\section{Conclusion}

We discussed the Standard Model with a modified scalar sector. Adding scalar particles which are 
singlets under the SM gauge group it was shown that hadron  colliders like the LHC may not be able 
to shed light on  the scalar sector of the theory by the most prominent Higgs signals.
 In the case of strong coupling between the SM Higgs scalar and those in the hidden sector the hopeful 
 Higgs signals could be  suppressed in such a way that detection is impossible.

The nonobservation of the SM Higgs boson 
 at the LHC or other hadron colliders could  indicate a slightly more 
complicated scalar sector. To be sure to find a Higgs boson one really has to build 
$e^+e^-$--colliders with adequate energies to produce  Higgs Bremsstrahlung by 
high energetic $Z$--bosons. But even then, due to the increased Higgs width, a missing
$p_T$ measurement should be more sensitive than in the SM case.     

\section*{Acknowledgements}
At this place, I would like to thank the organizers of the seminar ''Quarks 94''
for the wonderful atmosphere at the conference.

\end{document}